\documentstyle[12pt,notebook]{article}								 
\def\CellGroup{\bgroup}
\def\endCellGroup{\egroup}
\begin{document}
\thispagestyle{empty}
\vspace*{3cm}
\begin{center}
{\Large\bf 
Lepton transverse polarization in the $B \to D l \nu_l$ decay
due to the electromagnetic final state interaction.}
\end{center}

\vspace*{1cm}
\begin{center}
V.V. Braguta, A.E. Chalov, A.A. Likhoded$^{\dagger}$ 
\end{center}

\vspace*{1cm}
\begin{center}
{\it Institute for High Energy Physics, {\em{142280}}, \\
Protvino, Moscow region, Russia} \\

$^{\dagger}$ {\it Instituto de F\'isica Te\'orica - UNESP, \\
 Rua Pamplona, 145, 01405-900 S\~ao Paulo, SP, Brasil}\footnote{
On leave of absence from Institute for High Energy Physics, Protvino, 142284 Russia}\\
andre@ift.unesp.br
\end{center}

\normalsize
\vspace*{2cm}
\underline{\bf Abstract}

\vspace*{0.5cm}
\noindent
The effect of lepton transverse polarization in the 
$B^0 \to D^- l^+ \nu_l$, $B^+ \to \bar D^0 l^+ \nu_l$ decays ($l= \tau, \mu$)
is analyzed within the framework of Standard Model in the leading order of HQET.
It is shown that the non-zero transverse polarization appears due to the
electromagnetic final state interaction. The diagrams with intermediate 
$D,\; D^*$ - mesons contributing to the non-vanishing $P_T$ are considered.
Regarding only the contribution of these mesons,
values of the $\tau$-lepton transverse polarization, 
averaged over the physical region,  
in the $B^0 \to D^- \tau^+ \nu_l$ and $B^+ \to \bar D^0 \tau^+ \nu_l$ decays
are equal to $2.60 \cdot 10^{-3}$ and $-1.59 \cdot 10^{-3}$, correspondingly.
In the case of muon decay modes the values of $\langle P_T \rangle$ are equal to
$2.97 \cdot 10^{-4}$ and $ -6.79 \cdot 10^{-4}$.

\newpage

\section{Introduction}

\noindent
In spite of the remarkable phenomenological success of the Standard Model (SM)
the problem of the $CP$-violation mechanism still remains unexplained. 
In SM the $CP$-violation appears due to the complexity
of the CKM matrix; however, there is a set of models offering other
mechanisms of $CP$-violation. 
For instance, the Weinberg three Higgs  boson doublets model 
stands for one of the simplest
SM extension, where $CP$-violation appears due to the complex Higgs boson couplings to
fermions [1].The investigation of the $CP$-violation phenomenon 
will help us to understand its mechanism
and hence, to clarify one of the fundamental problems of elementary particle physics.

The experimental observables sensitive to $CP$-violating effects are,
for example, transverse lepton polarization in weak decays
and $T$-odd correlation. Muon transverse polarization in the  
$K^+\to\mu^+\nu\pi^0$, $K^+ \to \mu^+ \nu \gamma$ processes 
is the object of intensive study by many theoretical and experimental groups.
In some SM extensions a non-zero transverse muon polarization appears
already at the tree level [2,3].
The SM contribution to lepton transverse polarization is equal to zero in the leading
order, and this fact explains the smallness of the SM background. 
Final state interaction gives rise to non-zero $CP$-conserving contribution to $P_T$.
In the $K^+ \to \mu^+ \nu \gamma$  decay 
the lepton transverse polarization appears at the one-loop level and varies
in the range of $(0.0 - 1.1) \cdot 10^{-3}$ on the Dalitz plot.
The $P_T$ value averaged over the physical region   with the cut on photon energy
$E_\gamma \geq 20$ MeV' is equal to $4.76 \cdot 10^{-4}$ [2]. 
In the $K^+\to \mu^+ \nu \pi^0$ decay the muon transverse polarization is of order 
$\sim 10^{-6}$ [4,5], and therefore this decay
is rather effective to search for new physics effects. 
The measurement of muon transverse
polarization in this process is carried out by the KEK-E246 experiment, 
where the following result is obtained [6]:
\begin{equation}
P_T = -0.0042 + 0.0049(stat.) + 0.0009(syst.). 
\end{equation}
This experimental result does not allow to state that the value of $P_T$ is
stipulated by new physics effects. However,  
an increase of experimental accuracy is planned in the nearest future, 
which seems very promising from the point of $CP$-violation research.

Another experimental observable, suppressed in SM, is the $T$-odd correlation
in the charged kaon decays [7] (the distribution of the decay width 
over the kinematical variable, which is the mixing product of the
final particle momenta, for instance, $\vec p_{\pi} \cdot [\vec p_{\mu} \times \vec q]$
in the $K^+ \to \pi^0 \mu^+ \nu \gamma$ decay).
The small SM contribution to this observable can be explained by the same reason,
as the SM contribution to the lepton transverse polarization in $K_{l2\gamma}$ decay.
Here, new perspectives to search for $T$-odd contributions from new physiscs
are connected  with the OKA experiment [8], where it is planned to achieve 
$\sim 7 \cdot 10^5$ events for the $K_{l3\gamma}$ decay.

Aside from $K$-meson decays, it is possible to study the lepton transverse 
polarization in similar $B$-meson decays. 
It should be noted that the value of $P_T$ is especially sensitive
to $CP$-violating Higgs boson Yukawa couplings in these decays. 
Obviously, in the case of $B \to D^{(*)}  \tau \nu_{\tau}$,
the value of transverse polarization, due to the complexity of these couplings 
is $(m_b m_{\tau})/(m_s m_{\mu}) \sim 800$ times greater then $P_T$
in analogues $K \to \pi \mu \nu_\mu $ decay.

In [9,10,11] the effects of $CP$-violating transverse polarization of
leptons in the decays $B \to D(^*)\l \nu $ in various SM extensions
are analyzed. From these studies it follows that
the $\tau$-lepton transverse polarization can 
have the values $P_T\le 1$ in models with $CP$-violation in the Higgs sector [9,10],
and $P_T \le 0.26 $ in the leptoquark models  [11]. 
Thus, one can expect that the value of transverse 
polarization in various extensions of SM is rather large.
However, to estimate the impact of  new physics and perform the dedicated study
of the polarization phenomenon 
it is necessary to carry out the calculation of SM contribution to this observable.
In this paper we calculate the $CP$-conserving SM contribution to
$P_T$ in the $B^+ \to \bar D^0 l^+ \nu $ decays ($l=\tau \;, \mu$).
For simplicity, the calculations are carried out in the framework of
HQET in the leading approximation of $1/m_Q$ expansion.
It is shown below that the value of transverse polarization is not equal to zero if and
only if there is a non-zero phase shift of decay formfactors.
Electromagnetic final state interaction induces the non-zero phase shift
at the one-loop level and that, in turn, results in non-zero value of $P_T$.
In our calculations we take into account only the $(D, D^*)$ doublet contribution 
to the transverse polarization.

In the next section we discuss the matrix elements  contributing to the
polarization value. In section 3  the procedure of transverse polarization 
calculation is given.
The last section contains results and discussion.  

\section{Matrix elements}

\noindent
The general form for the  $\langle D(D^*)|V^\mu(A^\mu)|B \rangle$ matrix 
elements is as follows:
\begin{eqnarray}
\nonumber \langle D(k)| V^{\mu}| B(p) \rangle&=&f_+ (p^{\mu}+k^{\mu}) + f_- (p^{\mu}-k^{\mu})\;, \\ \nonumber
\langle D(k)| A^{\mu}| B(p)\rangle&=&0 \;,\\ \nonumber
\langle D^*(k, \epsilon)| V^{\mu}| B(p)\rangle&=& - i v e^{\mu \nu \alpha \beta} \epsilon^*_{\nu} 
k_{\alpha} p_{\beta} \;, \\ 
\langle D^* (k, \epsilon) | A^{\mu}| B(p) \rangle&=& a_1 (\epsilon^*)^{\mu} + a_2 (\epsilon^* p) p^{\mu}+
a_3 (\epsilon^* p) k^{\mu}\;, 
\end{eqnarray}
where $V^{\mu}=\bar b \gamma^{\mu} c$ and $A^{\mu}=\bar b \gamma^{\mu} \gamma_5 c$.
 
 The $\langle D(k)| A^{\mu}| B(p)\rangle$ matrix element is equal to zero, since
it is impossible to construct the axial vector composed of two momenta available.
In our calculations we use the following definition of the Levi-Civita tensor: 
$\epsilon^{0 1 2 3} = 1$.

Estimates of transverse polarization are carried out in the leading order of HQET,
i.e. under the assumption of $m_b, m_c \rightarrow \infty $. In this approximation 
the formfactors of the process are expressed in terms of the Isgur-Wise function 
$\xi (v v ')$ [12,13] and,
accordingly, expressions (2) can be rewritten as:  
\begin{eqnarray}
\nonumber
\langle D(k)| V^{\mu}| B(p)\rangle&=& \frac {\xi ( \omega )} {\sqrt {m_D m_B}}
(m_D p^{\mu} +m_B k^{\mu} )\;,  \\ \nonumber
\langle D(k)| A^{\mu}| B(p)\rangle&=&0 \;,\\ \nonumber
\langle D^*(k, \epsilon)| V^{\mu}| B(p)\rangle&=& - i  \frac {\xi ( \omega )} {\sqrt {m_D m_B}}
e^{\mu \nu \alpha \beta} \epsilon^*_{\nu}  k_{\alpha} p_{\beta} \;,\\ 
\langle D^*(k, \epsilon)| A^{\mu}| B(p)\rangle&=& \frac {\xi ( \omega )} {\sqrt {m_D m_B}}
( (m_B m_D+ pk ) \epsilon^{* \mu} - (\epsilon^* p) k^{\mu})\;,
\end{eqnarray}
where $\omega = (p k) / (m_D m_B)$.
Except for the given matrix elements, it is necessary to take into account the
matrix elements of the vector current between $D $ and $D $ ($D ^ * $ and $D $) states.
In the framework of HQET the formfactors of these matrix elements are also expressed 
through Isgur-Wise function:   
\begin{eqnarray}
\nonumber
\langle D(p')| \bar c \gamma^{\mu} c| D(k)\rangle= {\xi ( \omega' )} 
( p'^{\mu}+k^{\mu} )\;,  \\ 
\langle D(p')| \bar c \gamma^{\mu} c | D^*( k, \epsilon)\rangle= - i 
\frac {\xi ( \omega' )} {m_D}
e^{\mu \nu \alpha \beta} \epsilon_{\nu}  p'_{\alpha} k_{\beta}\;, 
\end{eqnarray}
where $\omega' = (k p')/(m_D m_B)$.
The $D$ and $D^*$ mass difference is neglected in Eqs. (3) and (4) as it
does not contribute to these matrix elements in the leading order of HQET. 

The function $\xi$ is conventionally parameterized as:
\begin{equation}
\xi ( \omega ) = 1 - \rho^2 (\omega -1)\;.
\end{equation}
In numerical estimates we use the value $\rho^2=0.94 $ obtained in [14]
within the framework of potential quark model. This result is in good agreement with 
the experimental data [13]. Different values can not drastically change the 
numerical results, since the  kinematical area of the decay is quite narrow. 
In the $B \to D \tau \nu _ {\tau} $ decay, 
$\omega $ varies from $1$ to $(m_B^2 + m_D^2-m_\tau^2)/2 m_B m_D = 1.43$, and 
in the $B \to D \mu \nu _ {\mu} $ decay this value varies in the range of
 1-1.59.   

The $ \langle D | J_{em}^{\mu} | D^* \rangle $ and 
$ \langle D | J_{em}^{\mu} | D \rangle$ matrix elements, 
where $J_{em}^{\mu}$ is the electromagnetic current,  are also required. 
It is possible to write down this operator as the sum of heavy and light
quark components:  
\begin{equation}
J_{em}^{\mu}=J_{h}^{\mu}+J_{l}^{\mu}\;.
\end{equation}
The matrix elements of heavy component of 
this electromagnetic current are expressed through 
Isgur-Wise function (4):  
\begin{eqnarray}
\nonumber
\langle D(p')| J_{h}^{\mu} | D(k)\rangle= -q_c {\xi ( \omega' )}  ( p'^{\mu}+k^{\mu} 
) \;, \\ 
\langle D(p')| J_{h}^{\mu} | D^*( k, \epsilon)\rangle=  i q_c \frac {\xi ( \omega' 
)} {m_D}
e^{\mu \nu \alpha \beta} \epsilon_{\nu}  p'_{\alpha} k_{\beta}, 
\end{eqnarray}
where $q_c$ is $c$-quark charge. The matrix elements of $J_{l}^{\mu}$ have the form:
\begin{eqnarray}
\nonumber
\langle D(p')| J_{l}^{\mu} | D(k)\rangle= q_l f^1_l( q^2 )  ( p'^{\mu}+k^{\mu} )\;,  \\ 
\langle D(p')| J_{l}^{\mu} | D^*( k, \epsilon)\rangle=  i q_l \beta f^2_l ( q^2 )
e^{\mu \nu \alpha \beta} \epsilon_{\nu}  p'_{\alpha} k_{\beta}\;,
\end{eqnarray}
where $q_l$ is the light quark charge and $f^{(1,2)}_l (0)=1$. 
The constant $\beta$, evaluated in [15] is equal to $1.9 \mbox{ GeV}^{-1}$.
In our calculations we use $q^2$-dependence of $f^{i}_l ( q^2 )$
formfactors $(i=1,\; 2)$, obtained under the assumption of dominant contribution of
the $\omega$ and $\rho$-resonances to these formfactors [16].
Neglecting the $\omega$ and $ \rho$ mesons mass difference the expressions for
$f^{i}_l$ take the form:
 \begin{eqnarray}
f^{i}_l=\frac 1 {1- \frac {q^2} {m_{\rho}^2} }\;, ~~~ i=1,2
\end{eqnarray}
where $m_{\rho}$  is the $\rho$ meson mass.

\section{Lepton transverse polarization}

\noindent
The amplitude of $B \to D(D^*) l^+ \nu_l$ decay can be written as follows:
\begin{equation}
M = \frac {G_F} {\sqrt 2} V_{cb}^* \langle D| V^{\mu}-A^{\mu}|B\rangle  \bar u ( p_{\nu} ) 
(1+\gamma_5) \gamma_{\mu} v(p_l)\;, 
\end{equation}
where $G_F$ is the Fermi constant and $V_{cb}^*$ is the corresponding 
CKM matrix element.
Matrix elements $\langle D| V^{\mu}-A^{\mu}|B\rangle$ are discussed in 
the previous section. For the case of the $B \to D l \nu_l $
process it is convenient to introduce the following 
parameterization of the amplitude: 
\begin{equation}
M = \frac {G_F} {\sqrt 2} 
V_{c b}^*  \bar u(p_{\nu}) (1+\gamma_5) (C_1 \hat p +C_2) v(p_l)\;.
\end{equation}

It should be noted that Eq. (11) is the most general form of the decay amplitude.
The expressions for $C_1, C_2$ in the leading order of HQET can be written as follows: 
\begin{eqnarray}
C_1 &=&  \frac {\xi ( \omega) } {\sqrt {m_B m_D}}  (m_D+m_B)\;, \nonumber \\
C_2&=&   \frac {\xi ( \omega) } {\sqrt {m_B m_D}}  (m_B m_l)\;.
\end{eqnarray}
The partial width of the $B \to D l^+ \nu_l $ decay in the  
$B$-meson rest frame can be expressed as:
\begin{equation}
d \Gamma = \frac {\sum |M|^2} { 2 m_B } (2 \pi)^4 \delta ( p - p_D - p_l -p_{\nu})
\frac {d^3 p_D} {(2 \pi)^3 2 E_D } \frac {d^3 p_l} {(2 \pi)^3 2 E_l }
\frac {d^3 p_{\nu_l}} {(2 \pi)^3 2 E_{\nu_l} }\;,
\end{equation}
where summation over lepton and photon spin states is performed.

Introducing the unit vector along the muon spin direction in lepton rest frame,
$\bf \vec s $, where ${\bf \vec e}_i~(i=L,\: N,\: T)$ are the unit vectors
along the longitudinal, normal and transverse components of lepton polarization, 
one can write down the matrix element squared for  the transition
into the particular lepton polarization state in the following form:
\begin{equation}
|M|^2=\rho_{0}[1+(P_L {\bf \vec e}_L+P_N {\bf \vec e}_N+P_T {\bf \vec
e}_T)\cdot \bf \vec s]\;,
\end{equation}
where $\rho_{0}$ is the Dalitz plot probability density averaged over polarization states.
The  unit vectors ${\bf \vec e}_i$ can be expressed in terms of the
three-momenta of final particles:
\begin{equation}
{\bf \vec {e} }_L=\frac {\vec p_l} {|\vec p_l|}\;,~~~ 
{\bf \vec e}_N=\frac{\vec p_l \times (\vec p_D \times \vec p_l)}
{|\vec p_l \times (\vec p_D \times \vec p_l)|}\;,~~~
{\bf \vec e}_T=\frac{\vec p_D \times \vec p_l }{|\vec p_D \times \vec
p_l|}\: . 
\end{equation}
With such definition of ${\bf \vec e}_i$ vectors, $P_T, P_L$, and $P_N$ denote
transverse, longitudinal, and normal components of the muon polarization,
correspondingly.

The Dalitz plot probability density has the following form:
\begin{eqnarray}
\nonumber
\rho_0 = G_F^2 |V_{cb}|^2 ( 4 |C_1|^2 (p p_{\nu}) (p p_l)+2 |C_2|^2 (p_{\nu} p_l)-\\ 
-2 |C_1|^2 (p_{\nu} p_l) m_B^2 -4 m_l Re(C_2 C_1^*) (p p_{\nu}))\;.
\end{eqnarray}
The expression for transverse polarization can be written as follows:
\begin{equation}
P_T = \frac {\rho_T} {\rho_0}\;,
\end{equation}
where $\rho_T$  has the form
\begin{equation}
\rho_T = 4 G_F^2 |V_{cb}|^2 m_B 
{\mbox {Im} } (C_1 C_2^*) |\vec p_D \times \vec p_l|\;.
\end{equation}
Obviously, the lepton transverse polarization arises only in the case of nonzero
phase shift between $C_1$ and $C_2$ formfactors. 
At the tree level of SM these formfactors are real  and thus
the lepton transverse polarization in this case is equal to zero.
The non-vanishing $P_T$ arises due to the effect of final state interaction.
To calculate the imaginary parts of the formfactors one can use the
$S$-matrix unitarity as it has been done in [16] for 
the case of the $K^0 \to \pi \mu \nu$ decay.

The diagrams inducing the transverse lepton polarization 
are shown in Fig. 1. As it has been mentioned earlier, in our calculations we
take into account only the diagrams with intermediate $D, D^*$-mesons.
The contribution of these diagrams to the imaginary part of the 
decay amplitude can be written as follows:  
\begin{eqnarray}
\nonumber
{\mbox {Im}} M =  \frac {G_F} {\sqrt {2}} V_{cb}^* \frac {\alpha} {2 \pi}
\int \frac {d \rho} {k_{\gamma}^2} 
\bar u(p_{\nu}) (1+ \gamma_5) \gamma_{\sigma} 
(\hat k_l - m_l) \gamma_{\lambda} v(p_l) \cdot
\\ \cdot\sum_{n=D,D^*} \langle D| J_{em}^{\lambda} |n\rangle \langle n| 
V^{\sigma}-A^{\sigma} |B \rangle\;, 
\end{eqnarray}
where $k_l, k_D$ denotes the four-momenta of the 
intermediate $D$, $D^*$ mesons, correspondingly,
$k_{\gamma}^2 = (k_D-p_D)^2$ is the squared transferred momentum,
$d \rho$ is the two particle phase space.
Expressions for the
$\langle D| J_{em}^{\lambda} |n\rangle, \langle n| V^{\sigma}-A^{\sigma} |B \rangle$
matrix elements are given in the previous section.

It should be noted that $D$-meson contribution to the imaginary part of the decay 
amplitude comprises an infrared divergence,
but the latter doesn't affect  the value of transverse polarization. One may explain
this fact by the factorization of soft photon contribution, which, in turn, 
does not lead to non-zero phase difference of formfactors required for 
a non-vanishing transverse polarization.
So, we do not have to take into account divergent terms as it was done in [16].
As for the $D^*$-meson contribution it is free from infrared divergence,
since in this case the lower bound of transferred momentum is: 
\begin{eqnarray}
( -k_{ \gamma}^2 )_{min} = (m_{D^*}^2-m_D^2) \frac {m_l} {m_l + m_D}\;,
\end{eqnarray}
which is equal to 500 and 160 MeV for the $\tau$-lepton and muon cases, correspondingly.
Taking into account this numerical values one can conclude that it is necessary to
regard mass difference of 
$D$ and $D^*$-mesons when integrating over the phase space.

Evidently, there are contributions to transverse polarization coming from
excited $D$-meson states. But we suppose that this contribution can not change
the result dramatically, since the Isgur-Wise function defined in (5) 
is greater than that in the case of $B$-meson decay into excited $D$-meson. 
The Isgur-Wise function for transition into $(0^+,1^+)$ doublet,
$\tau_{1/2} (\omega)$, was calculated in [17]. The value of $\tau_{1/2} (1)$
obtained in this paper is equal to 0.24. One can see that
the heavy quark contribution to $P_T$ is proportional to second power 
of $\tau_{1/2} (1)$ and the light quark
contribution is linear in $\tau_{1/2} (1)$. Furthermore, the physical 
region, where the value of transverse polarization is not equal to zero 
is bounded by inequality $(p_l+p_D)^2 \geq (m_D+m_l)^2$. 
In the case of intermediate $D^*$-meson this 
restriction does not cut physical region significantly. Contrary to the case of
intermediate state with $D^*$-meson, the physical region for the $(0^+,1^+)$ case
is strongly confined which results in the reduction of average transverse polarization.

We do not give the analytical expressions for imaginary parts of formfactors and 
integrals entering the imaginary part of one loop decay amplitude for their complexity.

\section{Numerical results and discussion}

\noindent
It is convenient to use the following $x, y$ variables
\begin{eqnarray}
E_{D} = \frac {m_B} 2 x\:,\;\;\;\; E_{\gamma} = \frac {m_B} 2 y\;,
\end{eqnarray}
where $E_D$ and $E_\gamma$ are the $D$-meson and photon energies; 
$m_B$ is the $B$-meson mass.

The three-dimensional distributionss of lepton transverse polarization 
in the kinematical region  $(x, y)$ for the processes 
$B^0 \to D^- \tau \nu_{\mu}$, 
$B^0 \to D^- \mu \nu_{\tau}$, $B^+ \to \bar D^0 \tau \nu_{\mu}$, and 
$B^+ \to \bar D^0 \mu \nu_{\tau}$ are shown in Figs. 2, 4, 6, 8 correspondingly.
The contour lines for $P_T$ in these decays are shown in Figs. 3, 5, 7, 9.

Corresponding values of transverse polarization averaged over the physical 
region are as follows
\begin{center}
\begin{tabular}{|l|r|} \hline 
Decay & $<P_T>$ \\ \hline
$B^0 \to D^- \tau \nu_{\tau}$      & $2.60 \cdot 10^{-3}$ \\ \hline
$B^0 \to D^- \mu \nu_{\mu}$        & $2.97 \cdot 10^{-4}$ \\ \hline
$B^+ \to \bar D^0 \tau \nu_{\tau}$ & $-1.59 \cdot 10^{-3}$ \\ \hline
$B^+ \to \bar D^0 \mu \nu_{\mu}$   & $-6.79 \cdot 10^{-4}$ \\ \hline
\end{tabular}
\end{center}
As it was mentioned earlier, the lepton transverse polarization can be represented
as the sum of heavy and light quarks contributions.

Formfactors of heavy and light electromagnetic current defined in 
(7) and (8) have different scales. The heavy formfactors  scale 
is of the order of $\sim m_D$ and the light formfactors  scale  is 
smaller and it is of the order of $\sim m_{\rho}$.  
Due to this fact and $ \bar D^0 $-meson neutrality one can state that
the contribution to $P_T$ from the diagram with intermediate
$\bar D^0 $-meson in the $B^+ \to \bar D^0 l \nu $ decay  
is rather small in $k_{\gamma}^2 \ll m_{\rho}^2$ region and appreciably large 
in $k_{\gamma}^2 \gg m_{\rho}^2$ kinematical region.

The contribution to $P_T$ coming from the diagram 
with intermediate $D^*$-meson turns out to be larger than that one of the diagram
with intermediate $D $-meson due to $ \bar D^0 $-meson neutrality. 
This fact explains the sign difference of averaged 
transverse polarization in the $B^+ \to \bar D^0 l \nu $
and $B^0 \to D ^-l \nu $ decays. 

Finally, we would like to remark that the averaged values of $P_T$ in SM 
is quite small in comparison to some models predictions [9,10,11].
This allows one to conclude that $B$-meson decays provide appealing
possibility to search for new physics effects.

\section*{Acknowledgements}

\noindent
The authors thank Drs. V.V. Kiselev and A.K. Likhoded for fruitful discussion and 
valuable remarks. This woks is in part supported by the Russian Foundation for Basic Research,
grants 99-02-16558 and 00-15-96645; Russian Education Ministry, grant E00-3.3-62 
and CRDF MO-011-0.The work of A.A.Likhoded is partially supported by FAPESP under grant
01/0639-04.

\newpage

\normalsize
\vspace*{2cm}
\section*{References}

\vspace*{0.5cm}
\noindent

\begin{description}
\item[1.] S. Weinberg, {\em Phys. Rev. Lett.} {\bf 37} (1976), 651.
\item[2.] V. Braguta, A. Likhoded, A. Chalov {\bf hep-ph}/0105111
\item[3.] G. Belanger, C.Q. Geng, {\em Phys. Rev.} {\bf D44} (1991), 2789.
\item[4.] A.R. Zhitnitskii, Sov. J. Nucl. Phys. {\bf 31} (1980), 529.
\item[5.] V.P. Efrosinin et.al., Phys.Lett. {\bf B493} (2000), 293.
\item[6.] M. Abe et. al., Phys. Rev. Lett. {\bf 83} (1999), 4253 
\item[7.] V. Braguta, A. Likhoded, A. Chalov, {\em Phys.Rev.} {\bf D65} (2002), 054038.
\item[8.]  OKA Letter of Intent, (see also, V.F. Obraztsov, {\bf hep-ex}/0011033). 
\item[9.] R. Garisto, {\em Phys.Rev.} {\bf D51} (1995), 1107.
\item[10.] G.H. Wu, K. Kiers, J. N. Ng, {\em Phys.Rev.} {\bf D56} (1995), 5413.
\item[11.] J.P. Lee, {\bf hep-ph}/0111184.
\item[12.] N. Isgur, M.B. Wise, Phys. Lett. {\bf 232} (1989) 113; {\bf 237} (1990), 527.
\item[13.] M.  Neubert, Phys. Rep. {\bf 245 } (1994), 259.
\item[14.] V.V. Kiselev, Mod. Phys. Lett. {\bf v.10} {1995}, 1049.
\item[15.] R. Casalbuoni et. al. {\bf hep-ph}/9605342.
\item[16.] P.Colangelo, et.al., Phys. Lett. {\bf B316} (1993), 555.
\item[17.] L.D. Okun, I.B. Khriplovich Yad. Fiz.  {\bf v.6} {1967}, 821.
\item[18.] P.Colagelo, G. Nardulli and N. Paver Phys. Lett. {\bf B293 } (1992), 207.

\end{description}

\newpage

\section*{Figure captions}

\vspace*{1.cm}
\noindent
\begin{description}
\item[Fig. 1.] Feynman diagrams contributing to lepton transverse 
polarization in $B^0 \to D^- l \nu_l$ decay at one-loop level of SM. 
\item[Fig. 2.] The 3D plot for $\tau$-lepton transverse polarization
for the case of $B^0 \to D^- \tau \nu_{\tau}$ decay.
\item[Fig. 3.] The level lines for $\tau$-lepton transverse polarization 
for the case of $B^0 \to D^- \tau \nu_{\tau}$ decay.
\item[Fig. 4.]  The 3D plot for muon transverse polarization
for the case of  $B^0 \to D^- \mu \nu_{\mu}$ decay.
\item[Fig. 5.] The level lines for muon transverse polarization for the case of
$B^0 \to D^- \mu \nu_{\mu}$ decay.
\item[Fig. 6.] The 3D plot for $\tau$-lepton transverse polarization
for the case of  $B^+ \to \bar D^0 \tau \nu_{\tau}$ decay.
\item[Fig. 3.] The level lines for $\tau$-lepton transverse polarization for the 
case of $B^+ \to \bar D^0 \tau \nu_{\tau}$ decay.
\item[Fig. 4.]  The 3D plot for muon transverse polarization
for the case of $B^+ \to \bar D^0 \mu \nu_{\mu}$ decay.
\item[Fig. 5.] The level lines for muon transverse polarization for the case 
of  $B^+ \to \bar D^0 \mu \nu_{\mu}$ decay.
\end{description}

\newpage

\setlength{\unitlength}{1mm}
\begin{figure}[ph]
\bf
\begin{picture}(150, 200)
\put(40,150){\epsfxsize=8cm \epsfbox{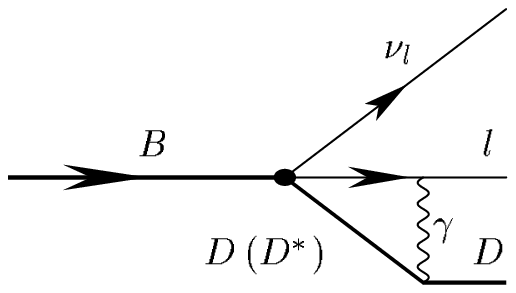}}
\put(70,145){Fig. 1}
\end{picture}

\newpage
\begin{picture}(150, 200)
\put(30,270){\epsfxsize=9cm \epsfbox{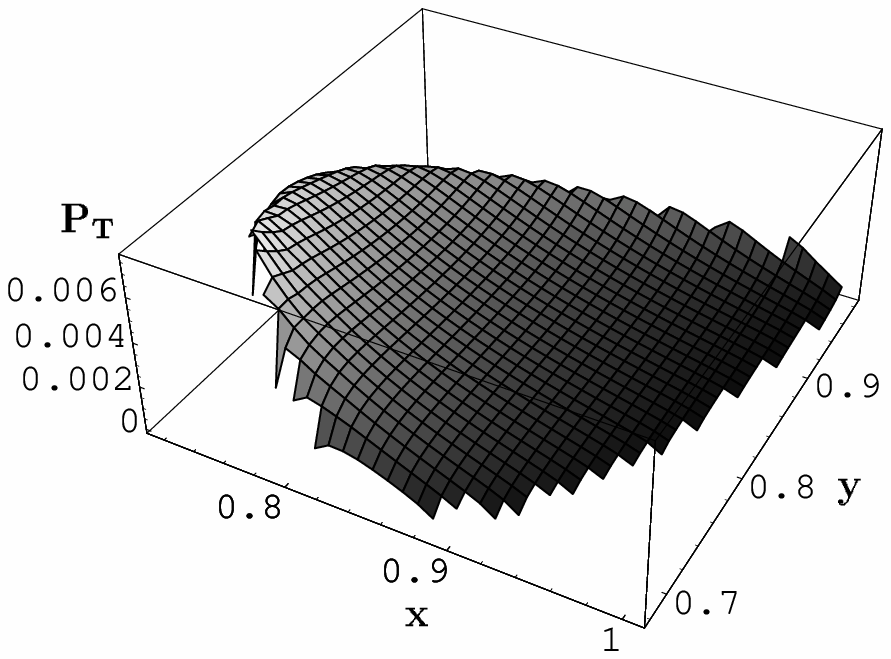}}
\put(70,265){Fig. 2}
\end{picture}

\begin{picture}(150, 200)
\put(25,390){\epsfxsize=10cm \epsfbox{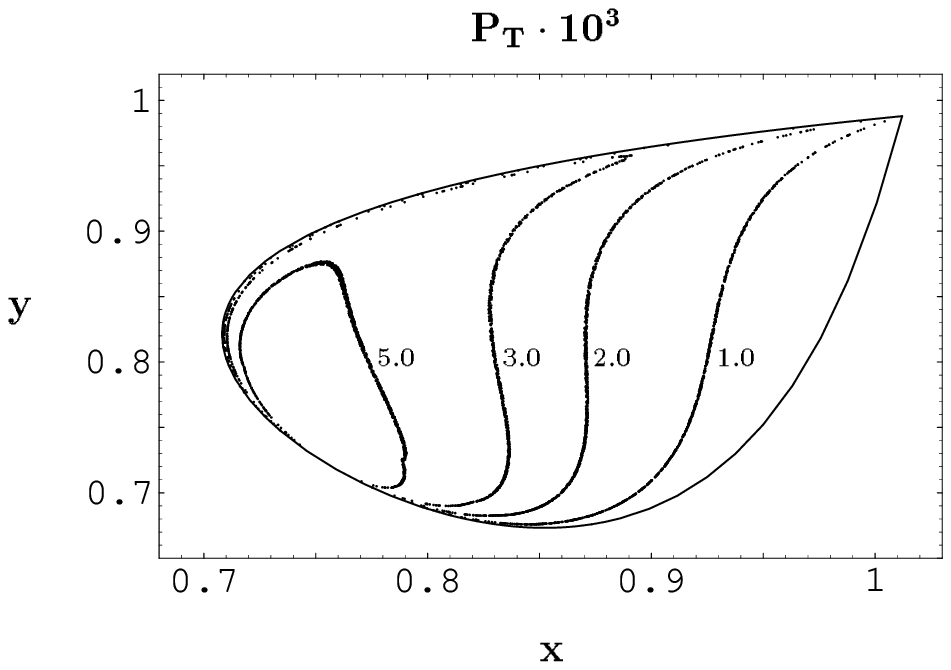}}
\put(70,380){Fig. 3}
\end{picture}

\end{figure}

\newpage

\setlength{\unitlength}{1mm}
\begin{figure}[ph]
\bf
\begin{picture}(150, 200)
\put(40,110){\epsfxsize=9cm \epsfbox{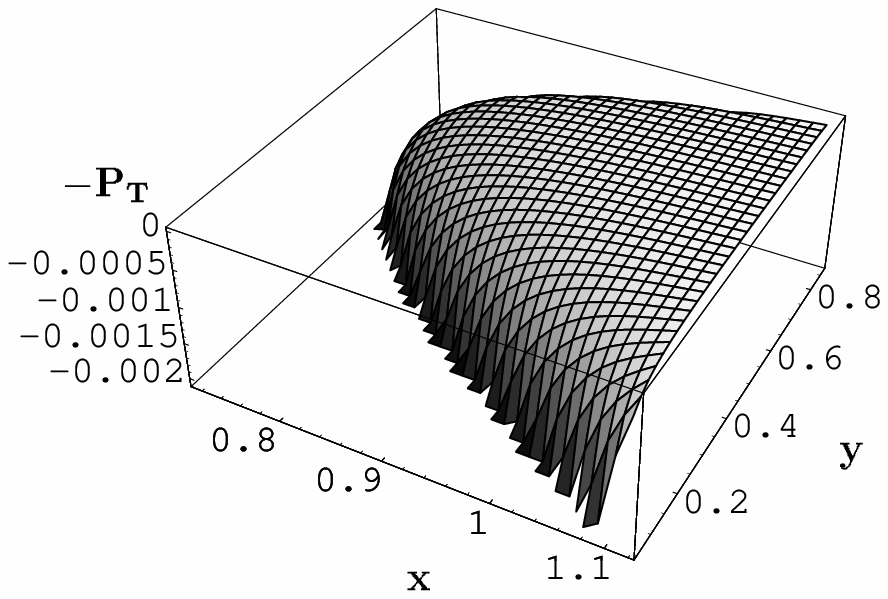}}
\put(70,105){Fig. 4}
\end{picture}

\vspace*{2cm}
\begin{picture}(150, 200)
\put(30,230){\epsfxsize=10cm \epsfbox{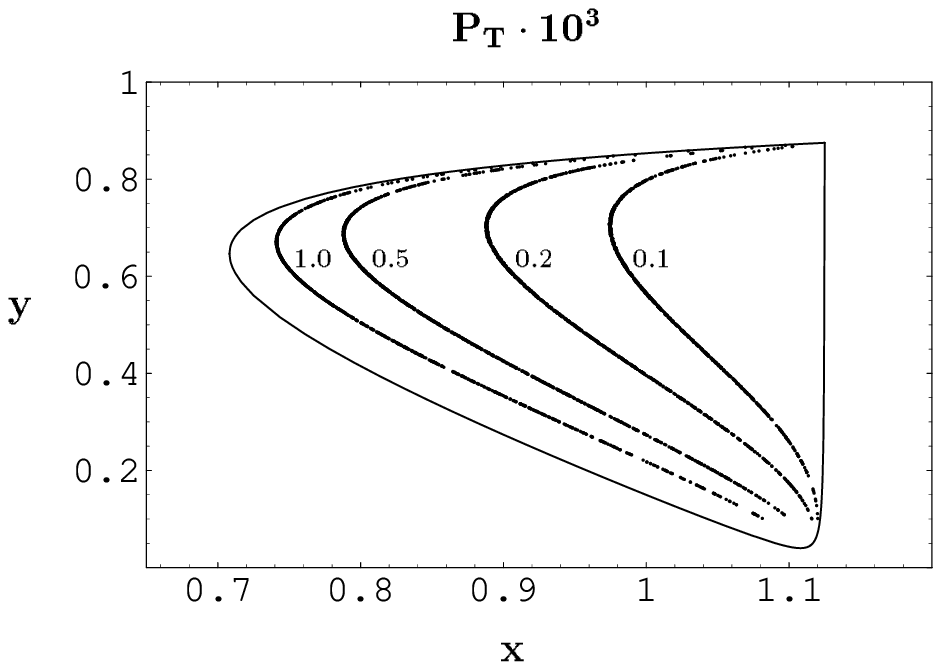}}
\put(70,225){Fig. 5}
\end{picture}

\end{figure}

\newpage

\setlength{\unitlength}{1mm}
\begin{figure}[ph]
\bf
\begin{picture}(150, 200)
\put(40,110){\epsfxsize=9cm \epsfbox{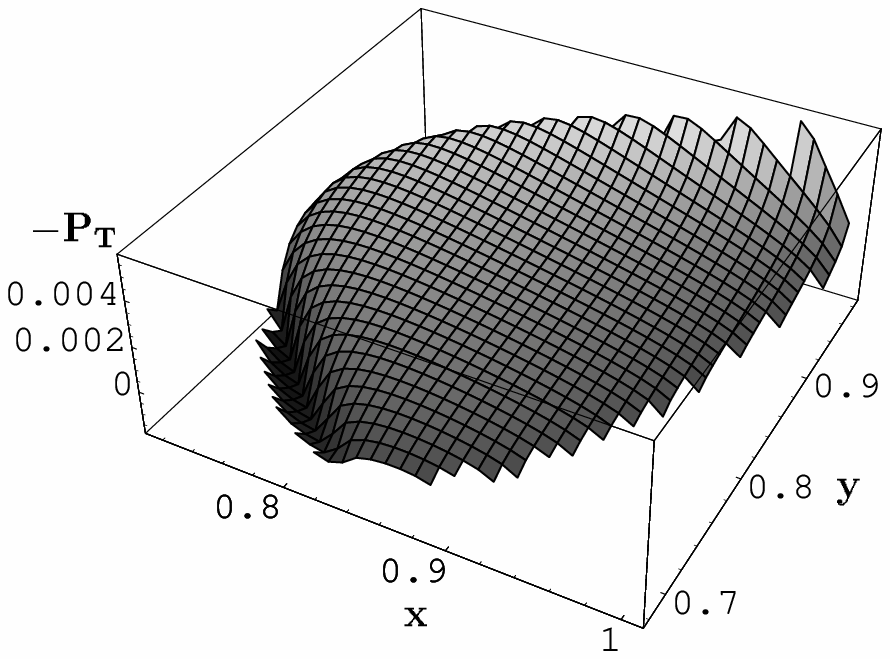}}
\put(70,105){Fig. 6}
\end{picture}

\vspace*{2cm}
\begin{picture}(150, 200)
\put(30,230){\epsfxsize=10cm \epsfbox{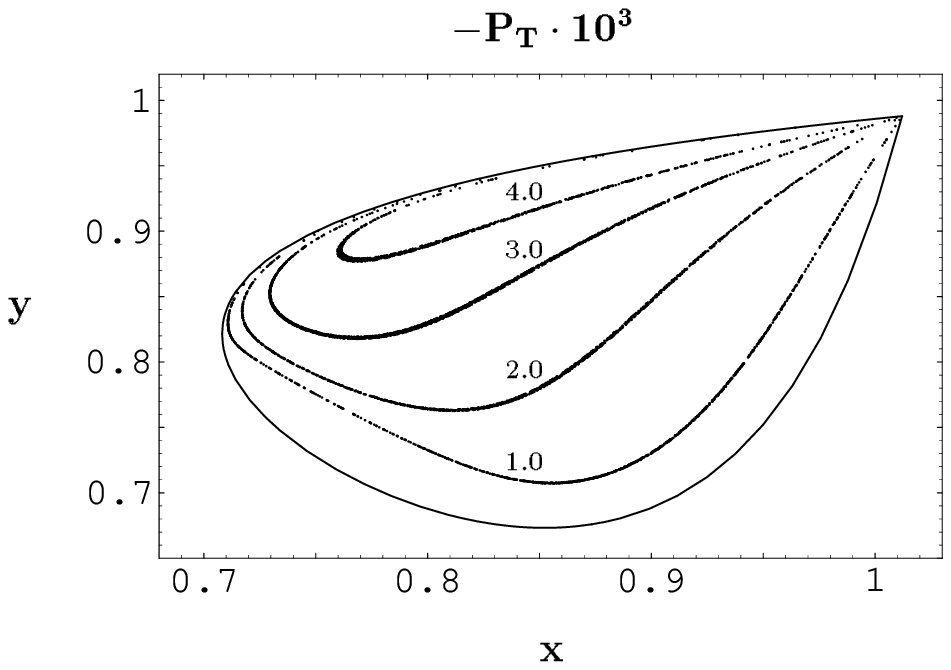}}
\put(70,225){Fig. 7}
\end{picture}

\end{figure}

\newpage

\setlength{\unitlength}{1mm}
\begin{figure}[ph]
\bf
\begin{picture}(150, 200)
\put(40,110){\epsfxsize=9cm \epsfbox{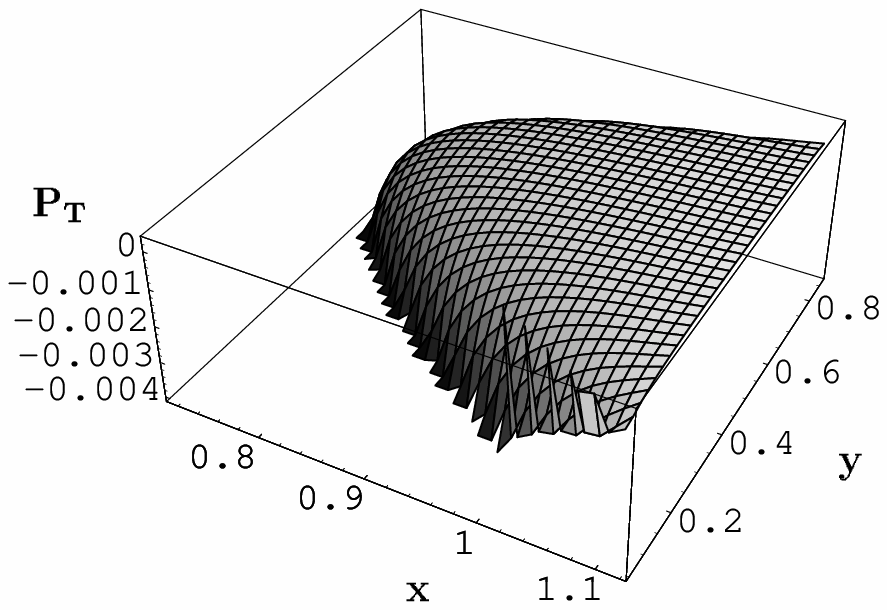}}
\put(70,105){Fig. 8}
\end{picture}

\vspace*{2cm}
\begin{picture}(150, 200)
\put(30,230){\epsfxsize=10cm \epsfbox{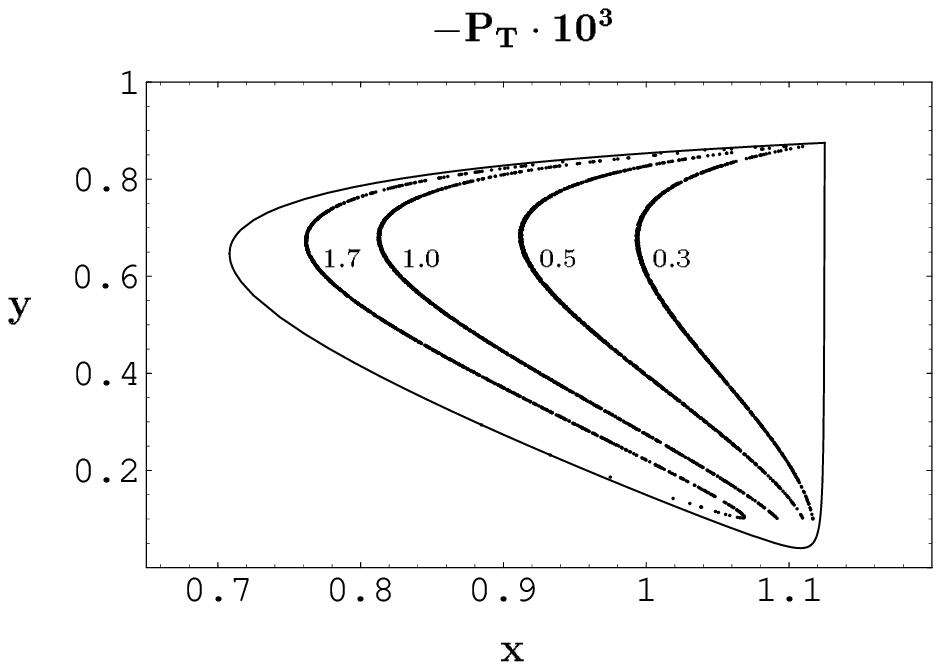}}
\put(70,225){Fig. 9}
\end{picture}

\end{figure}

\end{document}